\documentclass[12pt]{article}

\usepackage[margin=1in]{geometry}
\usepackage{amsmath,amssymb,amsfonts}
\usepackage{bm}
\usepackage{braket}
\usepackage{physics}
\usepackage{graphicx}
\usepackage{tikz}
\usetikzlibrary{arrows.meta, angles, quotes, calc} 

\usepackage{svg}
\svgsetup{
  inkscapeexe={"/Applications/Inkscape.app/Contents/MacOS/inkscape"}, 
  inkscapeversion=1
}

\usepackage{caption}
\usepackage{authblk}

\usepackage{float}

\usepackage[backend=biber,sorting=none,style=phys,articletitle=true,biblabel=brackets]{biblatex}

\usepackage{hyperref} 

\title{Geometric interpretation of the hyperfine Breit--Rabi solution}

\author{Lihong V. Wang}

\affil{Caltech Optical Imaging Laboratory\\
Andrew and Peggy Cherng Department of Medical Engineering\\
Department of Electrical Engineering\\
California Institute of Technology\\
1200 E. California Blvd., MC 138-78\\
Pasadena, CA 91125, USA\\
\texttt{LVW@caltech.edu}
}

\date{\today}

\begin{document}
\maketitle

\begin{abstract}
We present a geometric interpretation of the hyperfine Breit--Rabi eigenvalues and eigenvectors in alkali atoms after reformulating the standard solution into a compact form. In this picture, the nuclear magnetic moment has a polar angle fixed by the total projection quantum number. In contrast, the electron magnetic moment anti-aligns or aligns with an effective field formed by both the external magnetic flux density and the nuclear field, which simultaneously sets the mixing angle of the eigenvectors. This geometric view offers intuitive insight into the structure of the solutions.
\end{abstract}

\noindent\textbf{Keywords} hyperfine structure; Breit--Rabi; geometric interpretation; alkali atoms, quantum--classical correspondence

\section{Introduction}
The Breit--Rabi formula gives the dependence of hyperfine levels in atoms with a single valence electron on the magnetic flux density \cite{BreitRabi1931,Rabi1936,kusch1940radiofrequency}. The derivation proceeds by diagonalizing the hyperfine Hamiltonian. Here, we recast the solution in a geometric form that resolves the roles of the electron and nuclear magnetic moments and clarifies how the total angular momentum quantum number $F$ and the projection \(m_{F}\) quantize the relevant angles.

\begin{figure}[ht!]
\centering
\includegraphics[width=0.85\textwidth]{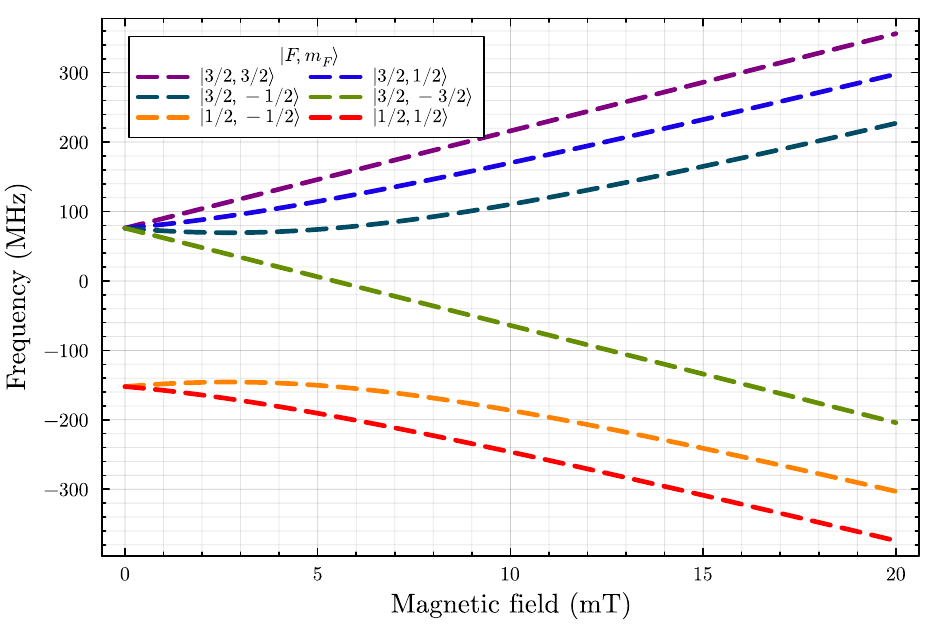}
\caption{Hyperfine Zeeman structure of \({}_3^{6}\mathrm{Li}\) with \(I=1\).}
\label{fig:Zeeman_Diagram_Li6}
\end{figure}

Figure \ref{fig:Zeeman_Diagram_Li6} illustrates a canonical hyperfine Zeeman diagram for \({}_3^{6}\mathrm{Li}\) with the nuclear spin quantum number \(I=1\). The Zeeman diagram displays the hyperfine energy levels of the system as a function of the applied magnetic flux density $B$. The top four levels constitute the upper manifold, corresponding to the total angular momentum quantum number $F=\frac{3}{2}$, while the lower two levels form the lower manifold with $F=\frac{1}{2}$. Within the upper manifold, the two straight lines represent the stretched states, whereas oblique states refer to all other Zeeman sublevels. Each manifold converges to a single frequency or energy as $B \to 0$. An essential feature of the diagram is the mirror symmetry between the upper and lower manifolds in the oblique states. For $I = \frac{1}{2}, \frac{3}{2}$, see Appendices \ref{app:H1} and \ref{app:K39}, respectively.

In the following, we first recapitulate the hyperfine Hamiltonian that governs the coupled electron–nuclear spin system, followed by the Breit–Rabi eigenstate solution in a finite field. We present a geometric interpretation of the Breit–Rabi solution, which provides intuitive insight into the underlying spin dynamics. Next, we geometrically represent the total angular momentum quantum numbers $F$ and $m_F$  in a vanishing field. A discussion follows, highlighting the physical implications of this picture. For completeness, we include in the appendices a recapitulated derivation of the Breit–Rabi solution and a reformulation of the associated mixing angle, as well as a reformulated Hamiltonian in Pauli form.

\section{Hyperfine Hamiltonian}

For an $s$-electron alkali atom in a static external magnetic flux density $\bm B$, the hyperfine Hamiltonian is
\begin{equation}
\hat{H}
=-\hat{\bm\mu}_{e}\cdot\bm B
-\hat{\bm\mu}_{n}\cdot\bm B
+A\,\hat{\bm S}\cdot\hat{\bm I},
\end{equation}
where $\hat{\bm S}$ and $\hat{\bm I}$ are the dimensionless electron and nuclear spin operators with spin quantum numbers $S$ and $I$, respectively; $\hat{\bm\mu}_{e}=\gamma_{e}\hbar \hat{\bm S}$ and $\hat{\bm\mu}_{n}=\gamma_{n} \hbar \hat{\bm I}$ are the electron and nuclear magnetic moment operators; $\gamma_{e}$ and $\gamma_{n}$ are the corresponding gyromagnetic ratios; $\hbar$ is the reduced Planck constant; and $A$ is the hyperfine coupling constant. Here, $S=\tfrac12$.

The total angular momentum operator is
\begin{equation}
\hat{\bm{F}} = \hat{\bm{S}} + \hat{\bm{I}},
\end{equation}
with $\hat{\bm{S}}$ and $\hat{\bm{I}}$ the electron and nuclear spin operators, respectively. The $z$-axis is taken along the direction of $\bm{B}$, i.e., \(\bm{z} \,\parallel\, \bm{B}.\) 

In the limit of an infinitesimal magnetic flux density ($B \to 0$), where $B = \abs{\bm B}$, the hyperfine Hamiltonian $\hat{H}$ commutes with the total angular momentum squared,
\begin{equation}
[\hat{H},\,\hat{\bm{F}}^{2}] \xrightarrow[B \to 0]{} 0.
\end{equation}
Thus, the magnitude quantum number $F$ is a good quantum number at \(B=0\) and an asymptotic good quantum number in this limit.  
We use the zero-field parent states $F$ as adiabatic labels for the two branches at fixed $m_F$.

For any values of $B$, the Hamiltonian commutes with the $z$-component of the total angular momentum,
\begin{equation}
[\hat{H},\,\hat{F}_{z}] = 0 \quad \forall\,B,
\label{eq:HFz_comm}
\end{equation}
which implies that the projection quantum number $m_{F}$ is always conserved.  

\section{Breit--Rabi eigenstate solution}
The eigenenergies for \(F=I\pm S\equiv F_\pm\) and projection \(m_{F}\) are derived for completeness in Appendix \ref{app:BR} \cite{BreitRabi1931,Rabi1936,kusch1940radiofrequency}:
\begin{equation}
E_\pm(m_F,B) =-\frac{h\Delta\nu}{2(2I+1)}-\mu_{N}g_{I}m_{F}B
\pm\frac{h\Delta\nu}{2}\sqrt{1+\frac{2m_{F}x}{I+S}+x^{2}}.
\end{equation}
where $\mu_{N}$ is the nuclear magneton and $g_{I}$ is the nuclear $g$-factor. The dimensionless field parameter \(x\) is defined by
\begin{equation}
x\equiv-\frac{\hbar(\gamma_e-\gamma_n)B}{A F_+}
=\frac{-g_e\mu_B+g_I\mu_N}{h\Delta\nu}\,B,
\label{eq:xDef-main}
\end{equation}
where $\mu_{B}$ is the Bohr magneton and $g_{e}<0$, in our sign convention, is the electron $g$-factor. Because \(-g_{e}\mu_{B}\) dominates, \(x>0\) for alkali atoms at \(B>0\). The zero-field hyperfine splitting energy determines the hyperfine coupling constant as follows:
\begin{equation}
h\Delta\nu=A(I+S),
\end{equation}
where $h$ is the Planck constant and $\Delta\nu$ is the frequency separation between the hyperfine levels at \(B=0\). 

The signed magnitudes of the electron and nuclear magnetic moments are
\begin{equation}
\mu_{e}=\gamma_{e}S\hbar < 0,\qquad
\mu_{n}=\gamma_{n}I\hbar.
\end{equation}
The electron gyromagnetic ratio is given by
\begin{equation}
\gamma_{e} = \frac{\mu_{B} g_{e}}{\hbar} < 0,
\end{equation}
where the negative sign reflects the negative charge of the electron.  The nuclear gyromagnetic ratio is
\begin{equation}
\gamma_{n} = \frac{\mu_{N} g_{I}}{\hbar}.
\label{eq:gamman_n}
\end{equation}
We consider only positive \(\gamma_{n}\) here because $\gamma_{n}>0$ for most alkali nuclei.

The original Breit--Rabi eigenvalue formula ensures that the mean energy of all levels is set to zero. An alternative option shifts the energy levels so that the mean of all oblique levels vanishes:
\begin{equation}
E_\pm(m_F,B) =-\mu_{N}g_{I}m_{F}B
\pm\frac{h\Delta\nu}{2}\sqrt{1+\frac{2m_{F}x}{I+S}+x^{2}}.
\label{eq:E_BR_sym}
\end{equation}

The stretched states ($m_{F}=\pm F_+$) correspond to $\bm{S}$ and $\bm{I}$ being parallel, producing maximal total angular momentum in the upper manifold. The oblique states ($|m_{F}|<F_+$) refer to all other Zeeman sublevels and correspond to different orientations between $\bm{S}$ and $\bm{I}$ for nonzero $B$, resulting in a total angular momentum vector $\bm{F}$ tilted relative to both spins. 

Since the projection quantum number $m_{F}$ is always conserved, for each fixed \(m_F\) and \(S=\tfrac12\), the relevant subspace is two-dimensional with the product basis
\begin{equation}
\ket{+}=\ket{m_S=+\tfrac12}\otimes\ket{m_I=m_F-\tfrac12},
\qquad
\ket{-}=\ket{m_S=-\tfrac12}\otimes\ket{m_I=m_F+\tfrac12}.
\label{eq:basis-main}
\end{equation}
The normalized eigenvectors corresponding to \(E_\pm\) are
\begin{equation}
\ket{\psi_\pm}=\cos\frac{\alpha}{2}\,\ket{\pm}\pm\sin\frac{\alpha}{2}\,\ket{\mp}.
\label{eq:eigvecs_main}
\end{equation}
The mixing angle \(\alpha\) is derived in Appendix \ref{app:alpha}:
\begin{equation}
\tan\alpha=\frac{B_n\sin\theta}{\eta_{\gamma}B+B_n\cos\theta},
\label{eq:tanalpha_main}
\end{equation}
where
\begin{equation}
\eta_{\gamma}\equiv1-\frac{\gamma_{n}}{\gamma_{e}}
=1+\mathcal{O}\!\left(\frac{1}{1836}\right)
\end{equation}
is introduced as a boost factor related to the gyromagnetic ratios for the electron Zeeman energy. Because the electron and nuclear gyromagnetic ratios differ by several orders of magnitude, the factor $\eta_{\gamma}$ exceeds unity only slightly. When the angle \(\alpha\) equals either \(0\) or $\pi$, we reach the stretched states, which are product states. In contrast, for other values of \(\alpha\), we obtain oblique states, which are entangled.

\section{Geometric interpretation of the Breit--Rabi solution}

In the zero-field limit, the eigenstates of the total angular momentum correspond to parallel $(F=I+S)$ or antiparallel $(F=I-S)$ coupling of the electron and nuclear magnetic moments. We introduce $\bm{B}_{n}$ to denote the effective magnetic flux density pseudovector at the electron produced by the nucleus; it is proportional to and parallel with the nuclear magnetic moment pseudovector $\bm{\mu}_{n}$ for an $s$-electron alkali atom via the Fermi-contact interaction \cite{bucher2000electron, wang2023multi}.   Therefore, the zero-field hyperfine splitting can be written as
\begin{equation}
h\Delta\nu
= \Delta (-\bm{\mu}_{e} \cdot \bm{B}_{n})
=-2\mu_{e}B_{n},
\label{eq:h_Delta_nu}
\end{equation}
Similarly, one can write \(h\Delta\nu=-2\mu_{n}B_{e}\). The quantity $\bm{B}_{e}$ denotes the effective magnetic flux density pseudovector at the nucleus produced by the electron and is proportional to and parallel with the electron magnetic moment pseudovector $\bm{\mu}_{e}$.  

The dimensionless field parameter can be rewritten as
\begin{equation}
x\equiv\frac{-\mu_{B}g_{e}+\mu_{N}g_{I}}{-2\mu_{e}}\frac{B}{B_{n}}
\equiv\frac{\eta_{\gamma}B}{B_{n}}.
\label{eq:x=alphB}
\end{equation}

Next, we reformulate the standard Breit--Rabi eigenvalue formula to facilitate interpretation. By inspecting Eq. \ref{eq:E_BR_sym} after substitutions for $h\Delta\nu$ (Eq. \ref{eq:h_Delta_nu}) and $x$ (Eq. \ref{eq:x=alphB}), we observe that the square root term arises naturally from the cosine law. Accordingly, we define the polar angle \(\theta\) such that
\begin{equation}
\cos\theta \equiv \frac{m_{F}}{I+S}.
\label{eq:cos_theta}
\end{equation}
Note that the denominator $I + S$ and the numerator \(m_{F}\in[-(I+S),I+S]\) are for both the upper and lower manifolds, and one should not use $F = I - S$ or $\sqrt{(I+S)(I+S+1)}$ \cite{feynman2005quantum} to determine $\theta$.  For now, this \(\theta\) is a bookkeeping angle determined by \(m_{F}\) and the fixed scale \(I+S\). We use it to cast energies in a compact cosine-law form.

Then, we also introduce a boost factor related to the angular momenta for the nuclear Zeeman energy
\begin{equation}
    \eta_{a}=1+\tfrac{S}{I}
\end{equation}
and an effective total field to be seen by the electron
\begin{equation}
\bm B_{t}=\eta_{\gamma}\bm B+\bm B_{n},\qquad
B_{t}=\sqrt{(\eta_{\gamma}B)^{2}+B_{n}^{2}+2\,(\eta_{\gamma}B)\,B_{n}\cos\theta}.
\label{eq:Bt}
\end{equation}

Consequently, Eq. \ref{eq:E_BR_sym}  takes the compact form
\begin{equation}
E_\pm 
=-{\mu}_n  \, \eta_{a} {B} \cos\theta
\mp \mu_e B_{t}
= -\bm{\mu}_n \cdot \eta_{a} \bm{B} - \bm{\mu}_e \cdot \bm{B}_t,
\label{eq:Epm}
\end{equation}
where the upper and lower signs correspond to the electron magnetic moment being antiparallel or parallel to \(\bm B_{t}\) as labeled by $F_\pm$. The first term on the right-hand side represents the nuclear Zeeman energy, where \(\eta_n B\) is the effective field due to hyperfine coupling. The second term encompasses both the electron Zeeman energy and the hyperfine coupling energy. Now, it is clear that the polar angle \(\theta\) determines that of the nuclear magnetic moment in the nuclear Zeeman contribution term, i.e., \(\theta_n = \theta\), regardless of the external field or the eigenstate. While the full vector expectation $\langle \hat{\bm I} \rangle$ varies with the mixing angle $\alpha$, this variation does not affect the term $-\mu_N g_I m_F B$.

The oblique states use \(m_{F}\) to set \(\theta\neq0\) or \(\pi\) , and the eigenenergies are split by \(\pm\mu_{e}B_{t}\). In contrast, the stretched states with \(m_{F}=\pm(I+S)\) have \(\theta=0\) or \(\pi\), and the eigenenergies reduce to
\begin{equation}
E_{s\pm}=\mp(\mu_{e}+\mu_{n})B-\mu_{e}B_{n}
= \mp \mu_n B -\mu_e (\pm B + B_n).
\end{equation}
Note that the boost factors have vanished. For the lower stretched state ($\theta = \pi$), we used $B_t = B_n - \eta_{\gamma} B$ at all field strengths when resolving the square root in Eq. \ref{eq:Bt}.

\begin{figure}[ht!]
    \centering
\begin{tikzpicture}[scale=1.0,>=stealth]


\coordinate (O1) at (-5, 0);

\coordinate (VlineTop) at (-5, 2.5);
\draw[-, thick, cyan] (O1) -- (VlineTop) node[midway, left] {$m_F$};

\coordinate (A) at (-5, 2*1.1);
\draw[->, thick] (O1) -- (A) node[below left] {$\eta_{a} \bm{B}$};

\def\spinAngle{30}
\def\spinLength{5}
\coordinate (SpinEnd) at ({-5 + \spinLength * cos(\spinAngle)}, {\spinLength * sin(\spinAngle)});
\draw[-, thick, cyan] (O1) -- (SpinEnd) node[midway, right, yshift=-9pt] {$I + S$};

\coordinate (munEnd) at ({-5 + \spinLength/3 * cos(\spinAngle)}, {\spinLength/3 * sin(\spinAngle)});
\draw[->, thick] (O1) -- (munEnd) node[below right] {$\bm{\mu}_n$};

\draw[dashed, thick] (VlineTop) -- (SpinEnd);

\draw [thick] pic["$\theta$", draw=black, angle radius=1cm] {angle=SpinEnd--O1--VlineTop};


\coordinate (O2) at (2.5, 0);

\coordinate (A) at (2.5, 2);
\draw[->, thick] (O2) -- (A) node[midway, right] {$\eta_{\gamma} \bm{B}$};

\coordinate (B) at ({2.5 - sqrt(3)}, -1);
\draw[->, thick] (B) -- (O2) node[midway, below right] {$\bm{B}_n \parallel \bm{\mu}_n$};

\draw[->, thick, dashed] (B) -- (A) node[midway, left] {$\bm{B}_t$};

\def\scaler{0.8}
\coordinate (Aorig) at ({2.5-1},2);
\coordinate (Borig) at ({2.5-sqrt(3)-1},-1);
\coordinate (A2) at ($ (Borig)!\scaler!(Aorig) $);
\coordinate (B2) at (Borig);

\coordinate (MidBA) at ($0.5*(B2) + 0.5*(A2)$);

\draw[<-, thick, blue] (B2) -- (MidBA) node[midway, left] {$\bm{\mu}_e+$};

\draw[<-, thick, red] (A2) -- (MidBA) node[midway, left] {$\bm{\mu}_e-$};

\draw [thick] pic[draw=black, angle radius=0.6cm, angle eccentricity=1.5]
{angle=A--O2--B};
\node at (3.2, 0.2) {$\pi - \theta$};

\draw [thick]
  pic["$\alpha$", draw=black, angle radius=0.6cm, angle eccentricity=1.3]
  {angle=B--A--O2};

\end{tikzpicture}

    \caption{Geometric interpretation of the exact Breit--Rabi eigenenergies for a given $m_F$, where $\theta$ is set by $m_F$ via \(\cos\theta \equiv \tfrac{m_{F}}{I+S}\) (Eq.~\eqref{eq:cos_theta}) and $\alpha$ is the polar angle of $\bm{B}_t$ via Eq.~\eqref{eq:tanalpha_main}. The left triangle shows that the nuclear magnetic moment \(\bm{\mu}_n\) has polar angle \(\theta_n = \theta\), independent of the external field magnitude or the eigenstates $\ket{\psi_+}$ and $\ket{\psi_-}$ for a given $m_F$. The right triangle shows the two allowed orientations of the electron magnetic moment \(\bm{\mu}_e\) relative to \(\bm{B}_t\), pointing either against ($+$) or along ($-$) \(\bm{B}_t\) for eigenstates $\ket{\psi_+}$ and $\ket{\psi_-}$, i.e., its polar angle \(\theta_e = \{\alpha,\pi+\alpha\}\) on the two-dimensional cross-section; it is the only quantity that varies between the two eigenstates. Alternatively, one can define the polar angle \(\theta\) to lie within the interval \([0, \pi]\) and the azimuthal angle \(\phi\) within \([0, 2\pi)\) to parameterize the two-sphere \(S^2\); then, \(\theta_e = \{\alpha,\pi+\alpha\}\) becomes \(\theta_e = \{\alpha,\pi-\alpha\}\) paired with \(\phi_e = \{0,\pi\}\).}
    \label{fig:Geo-Li6I1}
\end{figure}
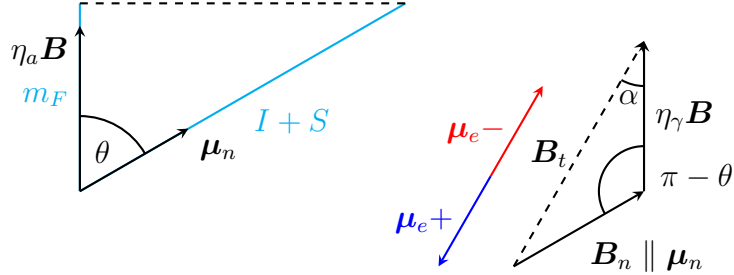

Figure \ref{fig:Geo-Li6I1} presents a plausible geometric interpretation of the exact Breit--Rabi eigenvalue formula, summarizing the above observations. It is essential to note the constraint $\bm{B}_{n} \parallel \bm{\mu}_{n}$. The electron magnetic moment \(\bm{\mu}_e\) is the only quantity that varies between the two eigenstates. 

The Hamiltonian block in the fixed-$m_F$ two-dimensional subspace can be reformulated into Pauli form (see Appendix \ref{app:H_sigma_form}):
\begin{equation}
H_{m_F}=-(\bm{\mu}_n \tau_0) \cdot (\eta_{a} \bm{B})- ({\mu}_e \bm\tau) \cdot \bm{B}_t,
\label{eq:H_mF_mu_main}
\end{equation}
where $\bm \tau$ is the effective Pauli operator vector (Eq. \ref{eq:tau-def}). It is illuminating to compare the Hamiltonian in this representation with the reformulated eigenvalues given in Eq. \ref{eq:Epm}, which we repeat below for convenience:
\begin{equation}
E_\pm 
=-{\mu}_n  \, \eta_{a} {B} \cos\theta
\mp \mu_e B_{t}
= -\bm{\mu}_n \cdot (\eta_{a} \bm{B}) - \bm{\mu}_e \cdot \bm{B}_t.
\label{eq:Epm-2}
\end{equation}
We conclude that the first term represents the nuclear Zeeman contribution, while the second term incorporates both the electron Zeeman and hyperfine-interaction energies. Note that the eigenstates are the \(\tau\)-spinors parallel or antiparallel to \(\bm b = -\mu_e \bm B_{t}\):
\begin{equation}
    \hat {\bm\tau}\cdot \bm n_{e}\,\ket{\psi_{\pm}} = \pm \ket{\psi_{\pm}}, 
\qquad 
\bm n_{e} = \frac{\bm b}{\lvert \bm b \rvert}.
\end{equation}

\begin{figure}[ht!]
    \centering

\begin{tikzpicture}[scale=1.5]
  \def\R{1.5}
  \def\alphadeg{32}
  \def\sep{4.2}

  \begin{scope}
    \draw[thick, red] (0,0) circle (\R);

    \draw[thick,->,>=Stealth] (0,-\R) -- (0,\R*1.2) node[above] {$\bm{B}$};

    \fill[red] (0,\R) circle (0.025);
    \node[above right, red] at (0,\R) {$\ket{+}$};
    \fill[red] (0,-\R) circle (0.025);
    \node[below right, red] at (0,-\R) {$\ket{-}$};

    \coordinate (PsiTipL) at ({\R*sin(\alphadeg)},{\R*cos(\alphadeg)});
    \draw[thick,->,>=Stealth, red] (0,0) -- (PsiTipL) node[midway, right, red] {$\ket{\psi_+}$};

    \coordinate (BtInL) at ({\R*sin(\alphadeg)},{\R*cos(\alphadeg)});
    \coordinate (BtOutL) at ({1.2*\R*sin(\alphadeg)},{1.2*\R*cos(\alphadeg)});
    \draw[thick,->,>=Stealth] (BtInL) -- (BtOutL) node[pos=0.95, right] {$\bm{B}_t$};

    \coordinate (MuTipL) at ({-\R*sin(\alphadeg)},{-\R*cos(\alphadeg)});
    \draw[thick,->,>=Stealth] (0,0) -- (MuTipL) node[midway, left] {$\bm{\mu}_e$};

    \draw[thick] (0,0) ++(0,0.45)
      arc[start angle=90, end angle={90-\alphadeg}, radius=0.45]
      node[pos=0.8, above=0pt] {$\alpha$};
  \end{scope}

  \begin{scope}[xshift=\sep cm]
    \draw[thick, red] (0,0) circle (\R);

    \draw[thick,->,>=Stealth] (0,-\R) -- (0,\R*1.2) node[above] {$\bm{B}$};

    \fill[red] (0,\R) circle (0.025);
    \node[above right, red] at (0,\R) {$\ket{+}$};
    \fill[red] (0,-\R) circle (0.025);
    \node[below right, red] at (0,-\R) {$\ket{-}$};

    \coordinate (MuTipR) at ({\R*sin(\alphadeg)},{\R*cos(\alphadeg)});
    \draw[thick,->,>=Stealth] (0,0) -- (MuTipR) node[midway, right] {$\bm{\mu}_e$};

    \coordinate (BtInR) at ({\R*sin(\alphadeg)},{\R*cos(\alphadeg)});
    \coordinate (BtOutR) at ({1.2*\R*sin(\alphadeg)},{1.2*\R*cos(\alphadeg)});
    \draw[thick,->,>=Stealth] (BtInR) -- (BtOutR) node[pos=0.95, right] {$\bm{B}_t$};

    \coordinate (PsiTipR) at ({-\R*sin(\alphadeg)},{-\R*cos(\alphadeg)});
    \draw[thick,->,>=Stealth, red] (0,0) -- (PsiTipR) node[midway, left, red] {$\ket{\psi_-}$};

    \draw[thick] (0,0) ++(0,0.45)
      arc[start angle=90, end angle={90-\alphadeg}, radius=0.45]
      node[pos=0.8, above=0pt] {$\alpha$};
  \end{scope}
\end{tikzpicture}

    \caption{For the two-level subspace, Bloch-sphere representation of the exact Breit--Rabi eigenstates of $\hat{H}$ at a given $m_F$ (in red) and the real-space quantities from Fig. \ref{fig:Geo-Li6I1} associated with the eigenvalues (in black). Note that \(\bm{\mu}_e\) here points either against or along \(\bm{B}_t\), i.e., its polar angle \(\theta_e = \{\alpha,\pi+\alpha\}\) on the two-dimensional cross-section. Note that \(\ket{\pm}=\ket{m_S=\pm\tfrac12}\otimes\ket{m_I=m_F\mp\tfrac12}\), \(H_{m_F}=-\bm{\mu}_n \tau_0 \cdot \eta_{a} \bm{B}- {\mu}_e \bm\tau \cdot \bm{B}_t\),  \(\ket{\psi_\pm(m_F,B)}=\cos\frac{\alpha}{2}\,\ket{\pm}\pm\sin\frac{\alpha}{2}\,\ket{\mp}
\), and \(E_\pm(m_F,B) 
=-{\mu}_n  \, \eta_{a} {B} \cos\theta
\mp \mu_e B_{t}
= -\bm{\mu}_n \cdot \eta_{a} \bm{B} - \bm{\mu}_e \cdot \bm{B}_t\).}
    \label{fig:Geo-Li6I1-psi}
\end{figure}
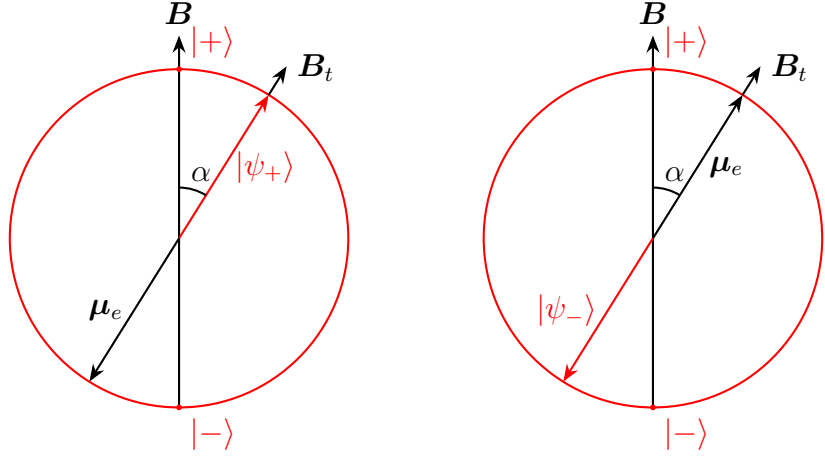

The effective field \(\bm B_t\) determines the mixing angle \(\alpha\), as shown in Appendix \ref{app:alpha}; it also determines the direction of the electron magnetic moment (see Appendix \ref{app:H_sigma_form}). Any pure state of a two-level system can be represented as a point on the Bloch sphere, with the two orthonormal basis states conventionally located at the north and south poles. Therefore, the eigenvectors at a given $m_F$ can be represented using the Bloch sphere  (Fig. \ref{fig:Geo-Li6I1-psi}); in real space, the electron magnetic moment pseudovector is either anti-aligned or aligned with \(\bm B_t\) (see Appendix \ref{app:H_sigma_form}). 

\section{Geometric interpretation at \(B \to 0\)}
\begin{figure}[h!]
    \centering

\begin{tikzpicture}[scale=0.5]
  \def\R{2}
  \def\r{0.8}
  \def\dx{7cm}     
  \def\dy{6cm}     
  \def\labelyA{3.2} 
  \def\labelyB{2.0} 

  \def\mS{0.5}
  \foreach \mF/\mFtext [count=\i] in 
    {1.5/{\tfrac{3}{2}}, 0.5/{\tfrac{1}{2}}, -0.5/{-\tfrac{1}{2}}, -1.5/{-\tfrac{3}{2}}} {
    \pgfmathsetmacro{\mI}{\mF-\mS}
    \pgfmathsetmacro{\cosval}{\mF/1.5}
    \pgfmathsetmacro{\cosval}{max(min(\cosval,1),-1)}
    \pgfmathsetmacro{\theta}{acos(\cosval)}
    \pgfmathsetmacro{\dxn}{\r*sin(\theta)}
    \pgfmathsetmacro{\dyn}{\r*cos(\theta)}
    \pgfmathsetmacro{\dxe}{-\R*sin(\theta)}
    \pgfmathsetmacro{\dye}{-\R*cos(\theta)}

    \begin{scope}[xshift={\i*\dx}]
      \ifnum\i<4
        \draw[->,gray] (0,0) -- (0,1) node[above] {$z$};
      \else
        \draw[->,gray] (0,0) -- (0,1) node[right] {$z$};
      \fi
      \draw[->,blue,very thick] (0,0) -- (\dxn,\dyn) node[right]  {$\bm{\mu}_n$};
      \draw[->,red,thick]  (0,0) -- (\dxe,\dye) node[left] {$\bm{\mu}_e$};
      \ifnum\i=1
        \node at (0,-\labelyA) {$F = \tfrac{3}{2},\, m_F = \mFtext$};
      \else
        \node at (0,-\labelyA) {$\mFtext$};
      \fi
    \end{scope}
  }

  \def\mS{-0.5}
  \foreach \mF/\mFtext [count=\i] in 
    {-0.5/{-\tfrac{1}{2}}, 0.5/{\tfrac{1}{2}}} {
    \pgfmathsetmacro{\mI}{\mF-\mS}
    \pgfmathsetmacro{\cosval}{\mF/1.5}
    \pgfmathsetmacro{\cosval}{max(min(\cosval,1),-1)}
    \pgfmathsetmacro{\theta}{acos(\cosval)}
    \pgfmathsetmacro{\dxn}{\r*sin(\theta)}
    \pgfmathsetmacro{\dyn}{\r*cos(\theta)}
    \pgfmathsetmacro{\dxe}{\R*sin(\theta)}
    \pgfmathsetmacro{\dye}{\R*cos(\theta)}

    \begin{scope}[xshift={(1+\i)*\dx}, yshift=-\dy]
      \draw[->,gray] (0,0) -- (0,1) node[right] {$z$};
      \draw[->,blue, very thick] (0,0) -- (\dxn,\dyn) node[below]  {$\bm{\mu}_n$};
      \draw[->,red,thick]  (0,0) -- (\dxe,\dye) node[above right] {$\bm{\mu}_e$};
      \ifnum\i=1
        \node at (0,-\labelyB) {$F = \tfrac{1}{2},\, m_F = \mFtext$};
      \else
        \node at (0,-\labelyB) {$\mFtext$};
      \fi
    \end{scope}
  }
\end{tikzpicture}

\caption{Geometric interpretation of the electron and nuclear magnetic moments for ${}^{6}_{3}\mathrm{Li}$ with $I=1$ in the limit $B \to 0$.  
The $\bm{z}$ axis is taken along the direction of the magnetic flux density $\bm{B}$. The polar angle of $\bm{\mu}_{n}$ is determined by Eq.~\ref{eq:cos_theta}, whereas the alignment of $\bm{\mu}_{e}$ with respect to $\bm{\mu}_{n}$ is determined by $F$. Due to the negative gyromagnetic ratio, the electron’s magnetic moment points opposite to its spin pseudovector.}
    \label{fig:FmF_geometry_Li6I1}
\end{figure}
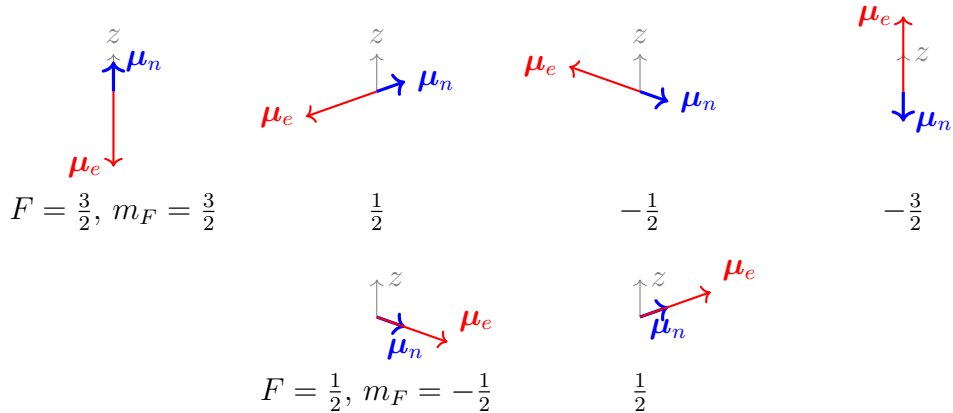

In the limit $B \to 0$, $F$ is an asymptotic good quantum number, and the allowed eigenstates are labeled by $F$ and $m_{F}$, and the corresponding eigenstates admit a more straightforward geometric interpretation. Figure~\ref{fig:FmF_geometry_Li6I1} illustrates the geometry of the total angular momentum $\bm{F}=\bm{S}+\bm{I}$ for ${}^{6}_{3}\mathrm{Li}$ with nuclear spin $I=1$ in the limit $B \to 0$. The polar angle of $\bm{\mu}_{n}$ is determined by Eq. \ref{eq:cos_theta}. The polar angle of $\bm{B}_{t}$
\begin{equation}
\lim_{x\to0}\alpha=\theta.
\label{eq:B=0-limit}
\end{equation}
The upper four configurations form the manifold with total angular momentum quantum number $F=\tfrac{3}{2}$, corresponding to the electron and nuclear spins $\bm{S}$ and $\bm{I}$ being aligned or their magnetic moments anti-aligned. In comparison, the lower two configurations constitute the manifold with $F=\tfrac{1}{2}$, which arises when $\bm{S}$ and $\bm{I}$ are oriented anti-aligned or their magnetic moments are aligned. Note that for the electron, the magnetic moment and the spin pseudovectors point in opposite directions due to the negative gyromagnetic ratio. Conversely, for the nucleus with a positive gyromagnetic ratio, the magnetic moment and the spin pseudovectors align in the same direction. See Appendices \ref{app:H1} and \ref{app:K39} for $I = \frac{1}{2}, \frac{3}{2}$, respectively. One may now introduce a field and vary it to rotate the electron magnetic moment $\bm{\mu}_e$ adiabatically according to the right panel in Fig. \ref{fig:Geo-Li6I1}.

\section{Discussion}

The geometric interpretation indicates that the nuclear magnetic moment $\bm{\mu}_n$ is determined by the polar angle $\theta$, independent of the external field $\bm{B}$, while the electron's magnetic moment aligns or anti-aligns with the effective field $\bm{B}_t$. As $\bm{B}$ increases in magnitude slowly, $\bm{B}_t$ changes accordingly, and the electron's magnetic moment $\bm{\mu}_e$ follows this field adiabatically. Stated differently, as the external field changes, $\bm{B}_t$ changes, and $\bm{\mu}_e$ follows it adiabatically, while $\bm{\mu}_n$ remains stable at a given $m_F$. This asymmetry likely arises because the nuclear gyromagnetic ratio is orders of magnitude smaller, making nuclear spin precession much slower, while the electron spin rapidly adapts. 

At $B=0$, we label the eigenstates by $\lvert F,m_F\rangle$ because $[\hat{H},\hat{\bm F}^{2}]=0$. For $B>0$, the hyperfine and Zeeman terms do not commute with $\hat{\bm F}^{\,2}$, so $F$ is not conserved. Nevertheless, $m_F$ remains conserved for all $B$ because of axial symmetry, and each fixed $m_F$ subspace is at most two-dimensional for $S=\tfrac12$. Turning on $B$ slowly, the adiabatic theorem guarantees that a state initially in $\lvert F_{\pm},m_F\rangle$ evolves into the corresponding instantaneous eigenstate $E_\pm(m_F,B)$ without interbranch transitions. It is therefore natural to keep the zero-field $F_\pm$ to label the two adiabatic branches for each $m_F$, which justifies calling $F$ an adiabatic label; \(F\) itself is not conserved for \(B>0\). In particular, the stretched branches carry unique zero-field parents and remain unmixed as product states at all $B$. In contrast, the oblique branches are the adiabatic continuations of the $F_{+}$ and $F_{-}$ parents, but they become mixtures of the two manifolds as entangled states as the field increases. 

Even for protium (\( {}^{1}_{1}\mathrm{H} \), $I=S$), the electron and nucleus behave asymmetrically, despite the apparent symmetry of the Hamiltonian. This asymmetry is likely rooted in the significant disparity between the electron and nuclear magnetic moments and their gyromagnetic ratios.

We will present the following discussions to encourage further investigation and dialogue. It is intended as a conceptual prompt rather than a definitive conclusion. Readers primarily interested in the main results may proceed to the next section.

The geometric interpretation, along with the above discussion, inspires the following adiabatic process to approximate the exact Breit--Rabi solution. Readers are referred to Fig. \ref{fig:Zeeman_Diagram_Li6}, Fig. \ref{fig:Zeeman_Diagram_H1}, and Fig. \ref{fig:Zeeman_Diagram_K39} for guidance, and should follow the field strengths indicated below.

(1) Electron spin in $\bm B_n$ without an external field (\(B=0\)).

In the absence of an external field, the electron spin collapses to an eigenstate defined by the nuclear field $\bm B_n$. The Zeeman interaction Hamiltonian is
\begin{equation}
\hat{H}_e^{(0)}=-\gamma_e \hat{\bm S}\cdot\bm B_n .
\end{equation}
Two hyperfine manifolds emerge as $F_+=I+S$ and $F_-=I-S$. 

(2) $\hat{\bm F}$ in an infinitesimally weak external field (\(B\to0\)).

In the weak field regime where $F$ remains an asymptotically good quantum number, the Zeeman interaction is well described by
\begin{equation}
\hat{H}_F=-\mu_B g_F\,\hat{\bm F}\cdot\bm B ,
\end{equation}
with $g_F$ the hyperfine Land\'e factor. The angle $\theta$ of the quantized direction satisfies
\begin{equation}
\cos\theta=\frac{m_F}{I+S} ,
\end{equation}
which agrees with Eq. \ref{eq:cos_theta}.

(3) Nuclear Zeeman energy in a finite-magnitude external field (\(B\neq0\)) without re-quantization.

To quantify the nuclear Zeeman contribution, we define
\begin{equation}
\hat{H}_n=-\hat{\bm \mu}_n\cdot\bm B=-\gamma_n \hat{\bm I}\cdot\bm B .
\end{equation}
However, we do not quantize the nuclear spin again; instead, we define its orientation by \(\theta_n = \theta\), determined under \(B\to0\), which also orients $\bm B_n \parallel \bm \mu_n$

(4) Electron spin in the total effective field $\bm B_t$.

When an external field is present, the electron spin is quantized along the total effective field \(\bm B_t\). The corresponding interaction Hamiltonian is
\begin{equation}
\hat{H}_e=-\hat{\bm \mu}_e\cdot\bm B_t=-\gamma_e \hat{\bm S}\cdot\bm B_t .
\end{equation}
This construction makes explicit that the electron senses both the applied field and the nuclear field. The slight deviation of \(\eta_{\gamma}\) in the expression of \(\bm B_t\) from unity due to hyperfine coupling might be related to the precessing \(\bm B_n\); thus, the natural quantization axis \(\bm B_t\) is not stationary.

(5) Total energy.

Collecting the contributions without double-counting the hyperfine coupling energy, we yield the following energy
\begin{equation}
E_\pm 
=-{\mu}_n  {B} \cos\theta
\mp \mu_e B_{t}
= -\bm{\mu}_n \cdot \bm{B} - \bm{\mu}_e \cdot \bm{B}_t.
\label{eq:Epm2}
\end{equation}
This equation reaches Eq. \ref{eq:Epm} if we set \(\eta_{a}  = 1\), which we found leads to only a minor error (typically \(10^{-4}\!-\!10^{-3}\)) due to the much smaller nuclear Zeeman contribution to the total energy.

Through the hyperfine coupling, the electron and nuclear spins mutually dress each other, giving rise to the joint eigenstates of the Breit--Rabi Hamiltonian (see Appendix \ref{app:BR_dressed}). Within each fixed-$m_F$ block, the nuclear Zeeman energy contribution boosted by $\eta_{a}$ through hyperfine coupling, \(-\,\eta_{a}\mu_{n}B\cos\theta\),  depends only on $\cos\theta = m_F/(I+S)$, which is common to both eigenstates. While $\alpha$ depends on $B$, $\theta$ does not. A possible geometric picture treats the $\bm{\mu}_n$ as aligned with the quantization direction set by $m_F$; in operator language, components transverse to $\hat F_z$ do not affect the eigenenergy. Because the total angular momentum projection operator \(\hat F_z=\hat{S}_{z}+\hat{I}_{z}\) determines the angle \(\theta\), the nuclear spin appears to follow the same projection and its Zeeman energy is boosted with \(\tfrac{I+S}{I}\). 

In the fixed-$m_F$ subspace
\begin{equation}
\mathcal H_{m_F}=\mathrm{span}\{\ket{+},\ket{-}\},
\end{equation}
using $\hat F_z=\hat S_z+\hat I_z$ yields
\begin{equation}
\hat F_z\ket{+}=\hbar m_F\ket{+},
\qquad
\hat F_z\ket{-}=\hbar m_F\ket{-},
\end{equation}
hence
\begin{equation}
\left.\hat F_z\right|_{\mathcal H_{m_F}}=\hbar m_F\,\mathbf 1.
\end{equation}
For any normalized $\ket{\psi}\in\mathcal H_{m_F}$
\begin{equation}
\bra{\psi}\hat F_z\ket{\psi}=\hbar m_F.
\end{equation}
Thus, $\hat F_z$ is proportional to the identity on $\mathcal H_{m_F}$, so any observable of the form $f(\hat F_z)$ cannot distinguish states within $\mathcal H_{m_F}$, consistent with the commutativity in Eq. \ref{eq:HFz_comm}. If the first Hamiltonian term depends only on $\hat F_z$, the dynamics and all expectation values coincide for all $\ket{\psi}\in\mathcal H_{m_F}$. In this restricted sense, the physics reduces effectively to a one-dimensional sector $\mathbb{C}^1$, although the underlying Hilbert space remains two-dimensional $\mathbb{C}^2$. Within the collapsed two-level subspace, the total angular momentum projection remains conserved, which permits treating the nuclear spin as a classical-like pseudovector. 

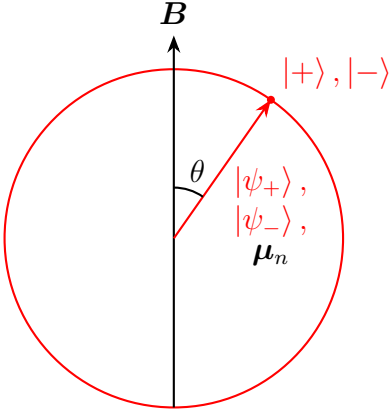
\begin{figure}[ht!]
    \centering
\begin{tikzpicture}[scale=1.5]
  \def\R{1.5}
  \def\thetadeg{35}

  \begin{scope}
    \draw[thick, red] (0,0) circle (\R);

    \draw[thick,->,>=Stealth] (0,-\R) -- (0,\R*1.2) node[above] {$\bm{B}$};

    \coordinate (PsiTip) at ({\R*sin(\thetadeg)},{\R*cos(\thetadeg)});
    \draw[thick,->,>=Stealth, red] (0,0) -- (PsiTip)
      node[midway, right=0.em, yshift=-1.5em, text=black]
      {\shortstack{\textcolor{red}{$\ket{\psi_+},$} \\
                   \textcolor{red}{$\ket{\psi_-},$} \\
                   $\bm{\mu}_n$}};

    \coordinate (KetTip) at ({\R*sin(\thetadeg)},{\R*cos(\thetadeg)});
    \node[above right, text=red] at (KetTip) {$\ket{+},\ket{-}$};
    \fill[red] (KetTip) circle (0.0353);

    \draw[thick] (0,0) ++(0,0.45)
      arc[start angle=90, end angle={90-\thetadeg}, radius=0.45]
      node[pos=0.8, above=0pt] {$\theta$};
  \end{scope}
\end{tikzpicture}

    \caption{For the collapsed two-level subspace, Bloch-sphere representation of the Breit--Rabi eigenstates of \(\hat{F}_{z}\) at a given $m_F$ (in red) and the real-space quantities including $\bm{\mu}_n$ in the dressed state from Fig. \ref{fig:Geo-Li6I1} associated with the eigenvalues (in black). Note that \(\cos\theta \equiv \tfrac{m_{F}}{I+S}, \ket{\pm}=\ket{m_S=\pm\tfrac12}\otimes\ket{m_I=m_F\mp\tfrac12}\), \(H_{m_F}=-\bm{\mu}_n \tau_0 \cdot \eta_{a} \bm{B}- {\mu}_e \bm\tau \cdot \bm{B}_t\),  \(\ket{\psi_\pm(m_F,B)}=\cos\frac{\alpha}{2}\,\ket{\pm}\pm\sin\frac{\alpha}{2}\,\ket{\mp}
\), and \(E_\pm(m_F,B) 
=-{\mu}_n  \, \eta_{a} {B} \cos\theta
\mp \mu_e B_{t}
= -\bm{\mu}_n \cdot \eta_{a} \bm{B} - \bm{\mu}_e \cdot \bm{B}_t\). }
    \label{fig:Geo-Li6I1-psi-n}
\end{figure}

Therefore, the eigenvectors of $\hat F_z$ at a given $m_F$ can be represented on a Bloch-sphere visualization of the fixed-$m_F$ subspace, emphasizing that the nuclear Zeeman contribution depends only on $\theta$; the sphere itself is still the two-level space spanned by $\{\ket{+},\ket{-}\}$ (Fig.~\ref{fig:Geo-Li6I1-psi-n}).
Here, the plus and minus basis vectors coincide in the same direction defined by a fixed polar angle $\theta$. Regardless of the mixing angle $\alpha$, the nuclear spin remains oriented in this direction. Strictly speaking, the full angular momentum operator $\hat{\mathbf F}$ determines the spin orientation, but within the fixed-$m_F$ subspace $\hat F_z$ already fixes the polar angle, and the transverse components provide no further distinction. Hence, it is sufficient to characterize the nuclear spin direction using $\hat F_z$ alone in this context.

\section{Conclusion}
We presented a geometric interpretation of the Breit--Rabi energies. The total angular momentum used to quantize the angle $\theta$ is taken as $I+S$ for both the upper and lower manifolds. It establishes a uniform quantization reference for both hyperfine manifolds. 

The quantized direction of the nuclear magnetic moment $\bm{\mu}_n$ is set by the angle $\theta$ regardless of the external field strength or the eigenstates at a given \(m_F\), giving a scaled energy \(-\left(1+\frac{S}{I}\right) \mu_n B \cos\theta\). This expression reflects the effective coupling of the nuclear magnetic moment to the applied field, providing a simple geometric picture for the nuclear contribution to the total energy. 

For the electron magnetic moment $\bm{\mu}_e$, the quantization is taken against or along the total effective field \(\bm{B}_t = \eta_{\gamma} \bm{B} + \bm{B}_n\), where \(\eta_{\gamma} = 1 - \frac{\gamma_n}{\gamma_e} = 1 + \mathcal{O}\!\left(\frac{1}{1836}\right)\). The effective field also sets the mixing angle of the eigenvectors. The resulting energy term $\mp\mu_e B_t$ incorporates both the applied field and the internal field from the nucleus. As one varies the external field, the electron's magnetic moment follows the effective field adiabatically, consistent with larger level splittings and faster Larmor precession for the electron; adiabaticity holds when the field is ramped slowly compared to the instantaneous eigenenergy gap. The \(\frac{\gamma_n}{\gamma_e}\) factor in \(\eta_{\gamma}\) due to hyperfine coupling may be related to the precessing quantization axis \(\bm{B}_t\).

\section{Acknowledgments}
The author thanks Kelvin Titimbo for providing the Zeeman diagrams (Figures \ref{fig:Zeeman_Diagram_Li6}, \ref{fig:Zeeman_Diagram_H1}, and \ref{fig:Zeeman_Diagram_K39}) and thanks our team members---Kelvin Titimbo, Suleyman Kahraman, Xukun Lin, Qihang Liu, Mark Zhu, and Arthur Chang---for their helpful discussions.

\printbibliography

\clearpage
\appendix

\section{Zeeman diagram and geometry for \( {}^{1}_1\mathrm{H} \), $I = 1/2$ \label{app:H1}}

\begin{figure}[H]
\centering
\includegraphics[width=0.75\textwidth]{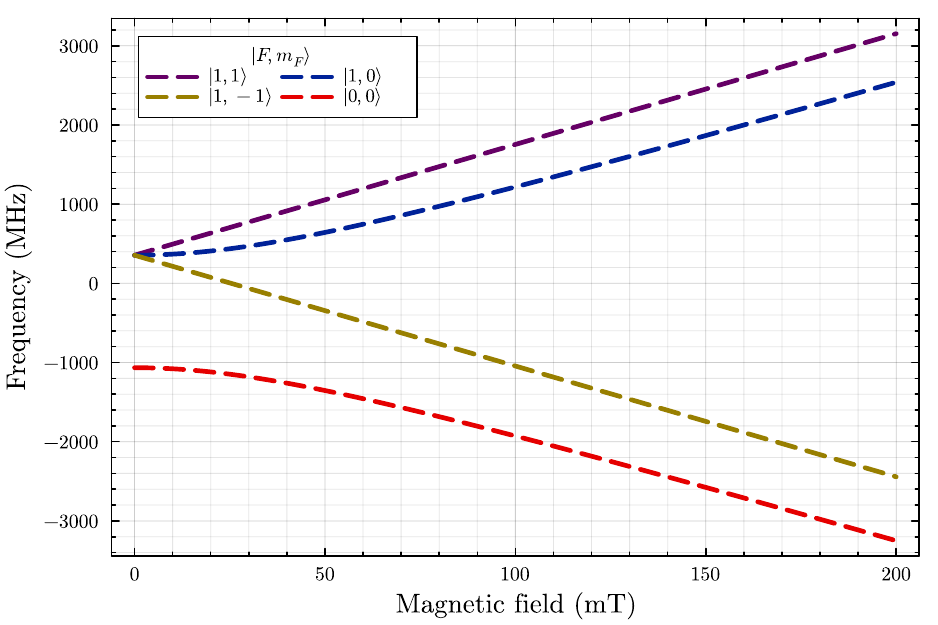}
\caption{Hyperfine Zeeman structure of \( {}^{1}_{1}\mathrm{H} \) with \(I=\frac{1}{2}\).}
\label{fig:Zeeman_Diagram_H1}
\end{figure}

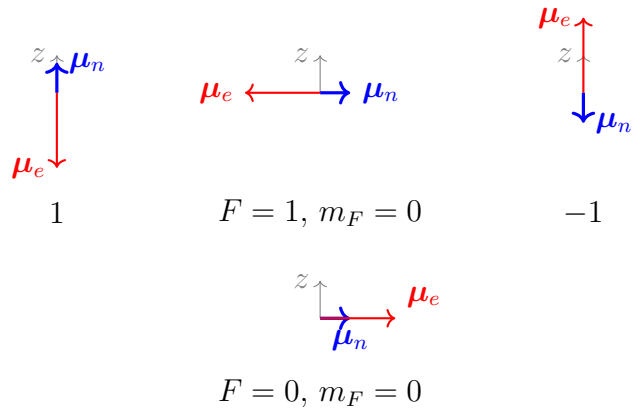
\begin{figure}[H]

\begin{center}
\begin{tikzpicture}[scale=0.5]
  \def\R{2}
  \def\r{0.8}
  \def\dx{7cm}     
  \def\dy{6cm}     
  \def\labelyA{3.2} 
  \def\labelyB{2.0} 

  \def\mS{0.5}
  \foreach \mF/\mFtext [count=\i] in 
    {1/{1}, 0/{0}, -1/{-1}} {
    \pgfmathsetmacro{\mI}{\mF-\mS}
    \pgfmathsetmacro{\cosval}{\mF/1.0}
    \pgfmathsetmacro{\cosval}{max(min(\cosval,1),-1)}
    \pgfmathsetmacro{\theta}{acos(\cosval)}
    \pgfmathsetmacro{\dxn}{\r*sin(\theta)}
    \pgfmathsetmacro{\dyn}{\r*cos(\theta)}
    \pgfmathsetmacro{\dxe}{-\R*sin(\theta)}
    \pgfmathsetmacro{\dye}{-\R*cos(\theta)}

    \begin{scope}[xshift={\i*\dx}]
      \ifnum\i<4
        \draw[->,gray] (0,0) -- (0,1) node[left] {$z$};
      \else
        \draw[->,gray] (0,0) -- (0,1) node[left] {$z$};
      \fi
      \draw[->,blue,very thick] (0,0) -- (\dxn,\dyn) node[right]  {$\bm{\mu}_n$};
      \draw[->,red,thick]  (0,0) -- (\dxe,\dye) node[left] {$\bm{\mu}_e$};
      \ifnum\i=2
        \node at (0,-\labelyA) {$F = 1,\, m_F = \mFtext$};
      \else
        \node at (0,-\labelyA) {$\mFtext$};
      \fi
    \end{scope}
  }

  \def\mS{-0.5}
  \def\mF{0}
  \pgfmathsetmacro{\mI}{\mF-\mS}
  \pgfmathsetmacro{\cosval}{\mF/1.0}
  \pgfmathsetmacro{\cosval}{max(min(\cosval,1),-1)}
  \pgfmathsetmacro{\theta}{acos(\cosval)}
  \pgfmathsetmacro{\dxn}{\r*sin(\theta)}
  \pgfmathsetmacro{\dyn}{\r*cos(\theta)}
  \pgfmathsetmacro{\dxe}{\R*sin(\theta)}
  \pgfmathsetmacro{\dye}{\R*cos(\theta)}

  \begin{scope}[xshift={2*\dx}, yshift=-\dy]
    \draw[->,gray] (0,0) -- (0,1) node[left] {$z$};
    \draw[->,blue, very thick] (0,0) -- (\dxn,\dyn) node[below]  {$\bm{\mu}_n$};
    \draw[->,red,thick]  (0,0) -- (\dxe,\dye) node[above right] {$\bm{\mu}_e$};
    \node at (0,-\labelyB) {$F = 0,\, m_F = 0$};
  \end{scope}

\end{tikzpicture}
\end{center}

\caption{Geometric interpretation of the electron and nuclear magnetic moments for ${}^{1}_1\mathrm{H}$ with $I=\frac{1}{2}$ in the limit $B \to 0$.  
The $\bm{z}$ axis is taken along the direction of the magnetic flux density $\bm{B}$.}
    \label{fig:FmF_geometry_H1Ihalf}
\end{figure}

\clearpage
\section{Zeeman diagram and geometry for \( {}^{39}_{19}\mathrm{K} \), $I = 3/2$ \label{app:K39}}

\begin{figure}[H]
\centering
\includegraphics[width=0.75\textwidth]{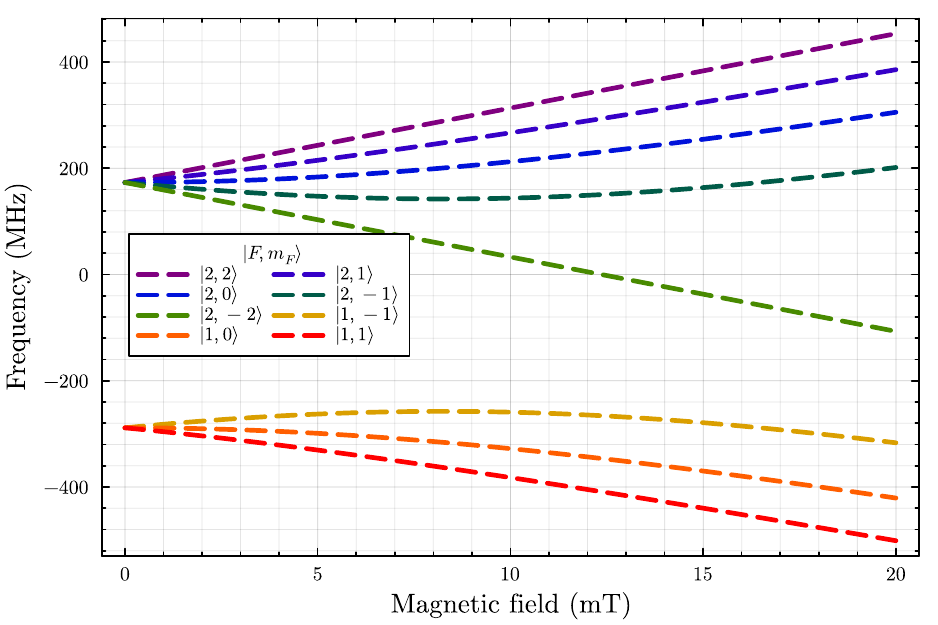}
\caption{Hyperfine Zeeman structure of \( {}^{39}_{19}\mathrm{K} \) with \(I=\frac{3}{2}\).}
\label{fig:Zeeman_Diagram_K39}
\end{figure}

\begin{figure}[H]

\begin{center}
\resizebox{\textwidth}{!}{

\begin{tikzpicture}[>=stealth,scale=1.2]
  \def\R{2}    
  \def\r{0.8}  
  \def\IS{2}   

  \foreach \mF/\mFtext [count=\i] in {
    2/{2}, 1/{1}, 0/{0}, -1/{-1}, -2/{-2}
  } {
    \pgfmathsetmacro{\cosval}{\mF/\IS}
    \pgfmathsetmacro{\cosval}{max(min(\cosval,1),-1)}
    \pgfmathsetmacro{\theta}{acos(\cosval)}
    \pgfmathsetmacro{\dxn}{\r*sin(\theta)}
    \pgfmathsetmacro{\dyn}{\r*cos(\theta)}
    \pgfmathsetmacro{\dxe}{-\R*sin(\theta)}
    \pgfmathsetmacro{\dye}{-\R*cos(\theta)}

    \begin{scope}[xshift={3*\i cm}]
      \draw[->,gray] (0,0) -- (0,1) node[right] {$z$};
      \draw[->,blue,very thick] (0,0) -- (\dxn,\dyn) node[right] {$\mu_n$};
      \coordinate (tip) at (0,0);
      \draw[->,red,thick]  (tip) -- ++(\dxe,\dye) node[left] {$\mu_e$};
      \node at (0,-2.4) {$F=2,\;m_F=\mFtext$};
    \end{scope}
  }

  \foreach \mF/\mFtext [count=\i] in {
    -1/{-1}, 0/{0}, 1/{1}
  } {
    \pgfmathsetmacro{\cosval}{\mF/\IS}
    \pgfmathsetmacro{\cosval}{max(min(\cosval,1),-1)}
    \pgfmathsetmacro{\theta}{acos(\cosval)}
    \pgfmathsetmacro{\dxn}{\r*sin(\theta)}
    \pgfmathsetmacro{\dyn}{\r*cos(\theta)}
    \pgfmathsetmacro{\dxe}{\R*sin(\theta)}
    \pgfmathsetmacro{\dye}{\R*cos(\theta)}

    \begin{scope}[xshift={(1+\i)*3cm},yshift=-4cm]
      \draw[->,gray] (0,0) -- (0,1) node[right] {$z$};
      \draw[->,blue,very thick] (0,0) -- (\dxn,\dyn) node[above] {$\mu_n$};
      \coordinate (tip) at (0,0);
      \draw[->,red,thick]  (tip) -- ++(\dxe,\dye) node[above] {$\mu_e$};
      \node at (0,-1.2) {$F=1,\;m_F=\mFtext$};
    \end{scope}
  }

\end{tikzpicture}

}

\end{center}

\caption{Geometric interpretation of the electron and nuclear magnetic moments for \( {}^{39}_{19}\mathrm{K} \) with $I = \frac{3}{2}$ in the limit $B \to 0$.  
The $\bm{z}$ axis is taken along the direction of the magnetic flux density $\bm{B}$.}
    \label{fig:FmF_geometry_K39I3halves}
\end{figure}

\clearpage

\section{Derivation of the Breit--Rabi solution \label{app:BR}}

\subsection{Hamiltonian and two-state basis}

We consider an alkali atom with electron spin \(S=\tfrac12\) and nuclear spin \(I\) in a static field of magnitude \(B\) along the laboratory \(z\) axis. The Zeeman and hyperfine Hamiltonian reads
\begin{equation}
\hat{H}
=-\hbar B\bigl(\gamma_e \hat S_z+\gamma_n \hat I_z\bigr)
+A\Bigl(\hat S_z \hat I_z+\tfrac12(\hat S_+\hat I_-+\hat S_-\hat I_+)\Bigr).
\label{eq:Hfull}
\end{equation}
We use \(\gamma_e=g_e\mu_B/\hbar\) and \(\gamma_n=g_I\mu_N/\hbar\) with \(g_e<0\). The subscripts $+$ and $-$ denote the ladder operators, with $+$ indicating raising and $-$ indicating lowering. The total magnetic quantum number \(m_F=m_S+m_I\) is conserved. 

For each fixed \(m_F\) and \(S=\tfrac12\), the relevant subspace is two-dimensional with the product basis
\begin{equation}
\ket{+}=\ket{m_S=+\tfrac12}\otimes\ket{m_I=m_F-\tfrac12},
\qquad
\ket{-}=\ket{m_S=-\tfrac12}\otimes\ket{m_I=m_F+\tfrac12}.
\label{eq:basis}
\end{equation}
We now restrict our attention to each subspace corresponding to a fixed \(m_F\).

\subsection{Matrix elements}

The diagonal matrix elements follow from \eqref{eq:Hfull} because \(\hat S_+\hat I_-+\hat S_-\hat I_+\) has no diagonal contribution in the product basis
\begin{equation}
H_{++}
=-\hbar B\Bigl(\gamma_e\tfrac12+\gamma_n(m_F-\tfrac12)\Bigr)
+A\Bigl(\tfrac12(m_F-\tfrac12)\Bigr),
\label{eq:Hpp}
\end{equation}
\begin{equation}
H_{--}
=+\hbar B\Bigl(\gamma_e\tfrac12-\gamma_n(m_F+\tfrac12)\Bigr)
-A\Bigl(\tfrac12(m_F+\tfrac12)\Bigr).
\label{eq:Hmm}
\end{equation}
The off-diagonal elements arise from the flip-flop term. Using the standard ladder matrix elements,
\begin{equation}
    \hat S_+\ket{S,m_S}=\sqrt{(S-m_S)(S+m_S+1)}\,\ket{S,m_S+1},
\end{equation}
\begin{equation}
    \hat I_-\ket{I,m_I}=\sqrt{(I+m_I)(I-m_I+1)}\,\ket{I,m_I-1},
\end{equation}
one obtains the equal off-diagonal elements as
\begin{align}
\chi &\equiv H_{+-} = H_{-+} \nonumber \\
&= \frac{A}{2}\,
\bra{S,+\tfrac12}\hat S_+\ket{S,-\tfrac12}\,
\bra{I,m_F-\tfrac12}\hat I_-\ket{I,m_F+\tfrac12} \nonumber \\
&= \frac{A}{2}\sqrt{(I+\tfrac12)^2-m_F^2}.
\label{eq:Delta}
\end{align}
For stretched states, \(\chi = 0\).

For convenience, we define the mean and half-difference of the diagonal elements as
\begin{equation}
M\equiv\frac{H_{++}+H_{--}}{2},
\qquad
d\equiv\frac{H_{++}-H_{--}}{2}.
\label{eq:MdeltaDef}
\end{equation}
With \eqref{eq:Hpp}--\eqref{eq:Hmm}, one finds
\begin{equation}
M=-\hbar\gamma_n m_F B-\frac{A}{4},
\qquad
d = -\tfrac{\hbar}{2}(\gamma_e-\gamma_n)B+\tfrac{A}{2}m_F.
\label{eq:MdeltaEval}
\end{equation}
Note that $d$ is related to the two Larmor frequencies.

\subsection{Eigenvectors and eigenvalues}

The eigenvalue equation satisfied by the eigenvectors \(\ket{\psi_\pm}\) and eigenvalues \(E_\pm\) is
\begin{equation}
\hat{H}_{m_F} \ket{\psi_\pm} = E_\pm \ket{\psi_\pm}.
\end{equation}
The Hamiltonian \(\hat{H}_{m_F}\) is expressed in the \(\{\ket{+}, \ket{-}\}\) basis as the Hamiltonian matrix
\begin{equation}
    H_{m_F}=\begin{pmatrix}H_{++}&\chi\\ \chi&H_{--}\end{pmatrix}.
\end{equation}

We define a mixing angle $\alpha$ via
\begin{equation}
\tan\alpha\equiv
\frac{\chi}{d}.
\label{eq:tanalpha_pairB}
\end{equation}

The normalized eigenvectors are
\begin{equation}
\ket{\psi_\pm}=\cos\frac{\alpha}{2}\,\ket{\pm}\pm\sin\frac{\alpha}{2}\,\ket{\mp}.
\label{eq:eigvecs}
\end{equation}
The corresponding eigenvalues are
\begin{equation}
E_{\pm}=M\pm\sqrt{d^2+\chi^2}
=\frac{H_{++}+H_{--}}{2}\pm\frac12\sqrt{(H_{++}-H_{--})^2+4\chi^2}.
\label{eq:Epm_general}
\end{equation}

Introduce \(F_+\equiv I+S\) and define the dimensionless field parameter \(x\) (Eq. \ref{eq:x=alphB}) by
\begin{equation}
x\equiv-\frac{\hbar(\gamma_e-\gamma_n)B}{A F_+}
=\frac{-g_e\mu_B+g_I\mu_N}{h\Delta\nu}\,B,
\qquad
h\Delta\nu=A F_+.
\label{eq:xDef}
\end{equation}

Using \eqref{eq:Delta}, \eqref{eq:MdeltaEval}, and \eqref{eq:xDef} in \eqref{eq:Epm_general}, and recalling \(\hbar\gamma_n =\mu_N g_I \) (Eq. \ref{eq:gamman_n}), one obtains the Breit--Rabi form:
\begin{equation}
E_\pm(m_F,B) =-\frac{h\Delta\nu}{2(2I+1)}-\mu_{N}g_{I}m_{F}B
\pm\frac{h\Delta\nu}{2}\sqrt{1+\frac{2m_{F}x}{I+S}+x^{2}}.
\label{eq:BreitRabi}
\end{equation}
The constant offset \(-h\Delta\nu/[2(2I+1)]\) can be dropped without affecting energy spacings or eigenvectors. 

\section{Reformulation of the mixing angle \label{app:alpha}}

Introduce the geometric angle \(\theta\) via Eq. \ref{eq:cos_theta}
\begin{equation}
\cos\theta=\frac{m_F}{F_+},
\qquad
\sin\theta=\sqrt{1-\left(\frac{m_F}{F_+}\right)^2}.
\label{eq:thetaDef}
\end{equation}
Using \eqref{eq:Delta}--\eqref{eq:MdeltaEval}, one gets
\begin{equation}
2 \delta = H_{++}-H_{--}=A F_+\bigl(x+\cos\theta\bigr),
\qquad
2\chi=A F_+\,\sin\theta.
\label{eq:detuning_in_theta}
\end{equation}
Substitution into \eqref{eq:tanalpha_pairB} gives the compact relation
\begin{equation}
\tan\alpha=\frac{\sin\theta}{x+\cos\theta}.
\label{eq:tanalpha_theta}
\end{equation}
Using Eq. \ref{eq:x=alphB}
\begin{equation}
x
\equiv
\frac{\eta_{\gamma}B}{B_{n}},
\end{equation}
we obtain
\begin{equation}
\tan\alpha=\frac{B_n\sin\theta}{\eta_{\gamma}B+B_n\cos\theta}.
\label{eq:tanalpha_theta_B}
\end{equation}
Therefore, the angle \(\alpha\) is the polar angle of \(\bm B_t\) measured from the \(+z\) axis.

Note that
\begin{equation}
\lim_{x\to0}\alpha=\theta.
\label{eq:branch}
\end{equation}
Substituting \(\alpha\to\theta\) into \eqref{eq:eigvecs} gives the Clebsch--Gordan limit:
\begin{equation}
\ket{\psi_+(m_F,0)}
=\cos\frac{\theta}{2}\,\ket{+}+\sin\frac{\theta}{2}\,\ket{-}
=\sqrt{\frac{F_++m_F}{2F_+}}\ \ket{+}
+\sqrt{\frac{F_+-m_F}{2F_+}}\ \ket{-},
\label{eq:CG_upper}
\end{equation}
\begin{equation}
\ket{\psi_-(m_F,0)}
=-\sin\frac{\theta}{2}\,\ket{+}+\cos\frac{\theta}{2}\,\ket{-}
=-\sqrt{\frac{F_+-m_F}{2F_+}}\ \ket{+}
+\sqrt{\frac{F_++m_F}{2F_+}}\ \ket{-}.
\label{eq:CG_lower}
\end{equation}

Equation \eqref{eq:tanalpha_theta} together with \eqref{eq:eigvecs} ensures that the eigenvectors are the electron spinors aligned and anti-aligned with the effective field \(\bm B_t\) of \eqref{eq:Bt}. The stretched states at \(m_F=\pm F_+\) satisfy \(\sin\theta=0\) and \(\chi=0\), hence \(\alpha\to 0\) or \(\pi\), and both \(\ket{\psi_+}\) and \(\ket{\psi_-}\) reduce to the unmixed product states at the edges, which confirms internal consistency.

\section{Reformulation of the Hamiltonian into Pauli form} \label{app:H_sigma_form}

In the fixed-$m_F$ two-dimensional subspace $\mathcal H_{m_F}=\mathrm{span}\{\ket{+},\ket{-}\}$, we define effective Pauli operators (including the identity)
\begin{align}
\hat{\tau}_x &= \ket{+}\!\bra{-}+\ket{-}\!\bra{+}, \nonumber\\
\hat{\tau}_y &= -i\bigl(\ket{+}\!\bra{-}-\ket{-}\!\bra{+}\bigr), \nonumber\\
\hat{\tau}_z &= \ket{+}\!\bra{+}-\ket{-}\!\bra{-}, \nonumber\\
\hat{\tau}_0 &= \ket{+}\!\bra{+}+\ket{-}\!\bra{-}.
\label{eq:tau-def}
\end{align}
The operators act only within this reduced space and should be clearly distinguished from the physical electron Pauli operators \( \hat \sigma_i \), acting on the physical electron spin space spanned by \(\left\{ \ket{m_s = +\tfrac{1}{2}}, \ket{m_s = -\tfrac{1}{2}} \right\} \).
The four effective Pauli matrices are
\begin{equation}
\tau_0=\begin{pmatrix}1&0\\[4pt]0&1\end{pmatrix},\quad
\tau_x=\begin{pmatrix}0&1\\[4pt]1&0\end{pmatrix},\quad
\tau_y=\begin{pmatrix}0&-i\\[4pt]i&0\end{pmatrix},\quad
\tau_z=\begin{pmatrix}1&0\\[4pt]0&-1\end{pmatrix}.
\end{equation}

Recall the Hamiltonian matrix in the constant-$m_F$ two-dimensional subspace 
\begin{equation}
    H_{m_F}=\begin{pmatrix}H_{++}&\chi\\ \chi&H_{--}\end{pmatrix}
    =\begin{pmatrix}M+d&\chi\\ \chi&M-d\end{pmatrix}.
\end{equation}
It is convenient to recast the Hamiltonian into Pauli form. Introducing the vector
\begin{equation}
\bm b=(\chi,0,d),
\end{equation}
one obtains
\begin{equation}
H_{m_F}=M\,\tau_0+\bm b\cdot \bm\tau.
\label{eq:H_mF_app}
\end{equation}
The vector $\bm b$ plays the role of a fictitious field, up to a constant factor, and turns out to be \(\bm B_t\). Its direction defines the sharp axis of the electron spin
\begin{equation}
\bm n_e=\frac{\bm b}{|\bm b|}=(\sin\alpha,0,\cos\alpha),
\qquad 
\tan\alpha=\frac{\chi}{d}.
\end{equation}

Diagonalization of the Hamiltonian is therefore immediate: the eigenvalues are $E_\pm=M\pm|\bm b|$, with eigenvectors $\ket{\psi_\pm}$ aligned or anti-aligned with $\bm n_e$. While the term $M$ gives the nuclear Zeeman energy, the term \(\pm|\bm b|\) gives the electron Zeeman energy and the hyperfine coupling energy.

From Eq. \ref{eq:MdeltaEval} and Eq. \ref{eq:Epm}, we find
\begin{equation}
    M = -\bm{\mu}_n \cdot \eta_{a} \bm{B},
\end{equation}
apart from a constant energy shift. Similarly, we have
\begin{equation}
    \bm b = - {\mu}_e \bm{B}_t .
\end{equation}
Therefore, we obtain
\begin{equation}
H_{m_F}=-(\bm{\mu}_n \tau_0) \cdot (\eta_{a} \bm{B})- ({\mu}_e \bm\tau) \cdot \bm{B}_t.
\label{eq:H_mF_mu_app}
\end{equation}

\section{Dressed-state analogy of the Breit--Rabi block}
\label{app:BR_dressed}
A dressed state is an eigenstate of the interacting system that includes both the matter degrees of freedom and the coupling that mixes the relevant bare states. The matter degrees of freedom are the bare hyperfine basis states $\ket{m_s,m_I}$, built from the electron spin $S$ and nuclear spin $I$. The interaction that mixes these bare states is the hyperfine flip–flop term, together with the Zeeman shifts that produce differential energies for the basis states. The resulting dressed state is therefore a coherent superposition of two bare configurations at fixed $m_F=m_s+m_I$.

In a fixed-$m_F$ subspace spanned by $\{\ket{+},\ket{-}\}$, the Hamiltonian can be written in Pauli form \(H_{m_F}=M\,\tau_0+d\,\tau_z+\chi\,\tau_x\) with the effective field \(\bm b=(\chi,0,d), |b|=\sqrt{d^2+\chi^2}\) .

For comparison, a driven two-level system in the rotating frame under the rotating-wave approximation has
\begin{equation}
H_{\mathrm{RWA}}=\frac{\hbar}{2}\bigl(-\Delta\,\sigma_z+\Omega\,\sigma_x\bigr),
\qquad
\Omega_R=\sqrt{\Delta^2+\Omega^2}.
\label{eq:RWA_block}
\end{equation}
The following parameter map identifies the two problems
\begin{equation}
d=-\frac{\hbar}{2}\,\Delta,\qquad
\chi=\frac{\hbar}{2}\,\Omega,\qquad
|b|=\frac{\hbar}{2}\,\Omega_R,
\label{eq:param_map}
\end{equation}
while the scalar offset $M\,\tau_0$ only shifts both energies equally and does not affect splittings within the constant-$m_F$ subspace.

The term $d\,\tau_z$ plays the role of a detuning inside the fixed-$m_F$ block. The term $\chi\,\tau_x$ is the coherent coupling that mixes the two bare states. In an alkali hyperfine manifold, $\chi$ originates from the flip-flop part of the contact hyperfine interaction $\tfrac{A}{2}(S_+I_-+S_-I_+)$. The vector $\bm b$ is the Bloch-vector representation of the effective field that sets the quantization axis for the spinor.

\end{document}